\documentclass{aa}
\usepackage{txfonts}
\usepackage{graphicx}
\usepackage{times}
\usepackage{natbib}
\usepackage{subfigure}
\usepackage{footnote}
\usepackage{lscape}
\usepackage{multirow}
\begin{document}
   \title{On the massive star content of the nearby dwarf irregular
Wolf-Rayet galaxy IC 4662\thanks{Based on observations made with ESO telescopes at the 
Paranal observatory under program ID 65.H-0705(A) and 
archival NASA/ESA Hubble Space Telescope datasets, 
obtained from the ESO/ST-ECF Science Archive Facility.}
}

\titlerunning{On the massive star content of the Wolf-Rayet galaxy 
IC~4662}
\authorrunning{Crowther \& Bibby}  

   \author{
          P. A. Crowther
          \and
          J. L. Bibby
          }

   \offprints{Paul.Crowther@sheffield.ac.uk}

   \institute{Department of Physics \& Astronomy, 
Hicks Building, University of Sheffield, Hounsfield Road, Sheffield, S3 7RH, UK
             }

   \date{Received 2 Feb 2009/Accepted 11 Mar 2009}

\abstract
{}
{We investigate the massive stellar content of the nearby  dwarf
irregular Wolf-Rayet galaxy IC~4662, and consider its global star 
forming properties in the context of other metal-poor galaxies, the
SMC, IC~10 and NGC~1569.}
{Very Large Telescope/FORS2 imaging and spectroscopy plus 
archival Hubble Space Telescope/ACS imaging datasets permit us to 
spatially identify the location, number and probable subtypes of 
Wolf-Rayet stars within this galaxy. We also investigate 
suggestions that a significant fraction of the ionizing photons of the two 
giant H\,{\sc ii} regions A1 and A2 lie deeply embedded within these 
regions.} 
{Wolf-Rayet stars are  associated with a number of sources within 
IC~4662-A1 and A2, plus a third compact H\,{\sc ii} region to the north 
west of A1 (A1-NW). 
Several sources appear to be isolated, single (or binary) luminous 
nitrogen sequence WR stars, while extended sources are clusters 
whose masses exceed the 
Orion Nebula Cluster by, at most, a factor of two. IC~4662 lacks 
optically visible young massive, compact clusters that are common in 
other nearby dwarf irregular galaxies. A comparison between radio and 
H$\alpha$-derived ionizing fluxes of A1 and A2 suggests that 
30--50\% of their total Lyman continuum fluxes lie deeply embedded within 
these regions. 
}
{The star formation surface density of IC~4662 is insufficient for this 
galaxy to qualify as a starburst galaxy, based upon its photometric radius, $R_{25}$.
If instead, we were to adopt the 
$V-$band  scale  length $R_{D}$ from Hunter \& Elmegreen, IC~4662 would comfortably
qualify as a starburst galaxy, since its star formation intensity would 
exceed $0.1 M_{\odot}$ yr$^{-1}$ kpc$^{-2}$.}


\maketitle

\section{Introduction}

Star-forming galaxies at high redshift, such as Lyman break galaxies,
possess metallicities comparable to the Small and Large Magellanic Clouds
(SMC, LMC), although typically with star formation surface densities
many orders of magnitude higher. Within the local universe, there are no 
direct analogues of massive, metal-deficient Lyman break galaxies, so one
must rely on lower mass, dwarf irregular galaxies to study 
starburst activity at low metallicity.

Dwarf, metal-poor starburst galaxies such as NGC~1569
\citep{Buckalew2006} have been studied in great
detail, but other examples at similar distances have largely been
neglected.  IC~4662 (IBm) is one such case. It has a SMC-like
metallicity of $\log$(O/H)+12 $\sim8.1$, possesses two giant H\,{\sc ii}
regions (A1 and A2) which dominate its visual appearance
\citep{Heydari1990}, and lies at a distance of 2.44 Mpc
\citep{Karachentsev2006}.  These H\,{\sc ii} regions host Wolf-Rayet
(WR) populations \citep{Rosa1986, Heydari1990, Richter1991},
indicating the presence of $\sim$5 Myr old stellar populations, such that
IC~4662 is one of the closest examples of a WR galaxy
\citep{Schaerer1999}. However the spatial location and number of WR
stars remains uncertain.  Wolf-Rayet stars in metal-poor galaxies such
as IC~4662 are believed to be the leading candidates for the
progenitors of Type Ic supernova and some 
long duration Gamma Ray Bursts (GRBs) \citep{Hammer2006, Modjaz2008}.

\citet{Johnson2003} have proposed that IC~4662 contains deeply embedded 
star forming  regions -- referred to as ultradense H\,{\sc ii} regions -- 
on the basis of radio observations. Such regions are typically very young and 
represent scaled-up versions of ultracompact H\,{\sc ii} regions. Such
regions become optically visible within $\leq 0.5-1$ Myr  according to 
Spitzer observations of H\,{\sc ii} regions within nearby 
star-forming galaxies \citep{Prescott2007}. If A1 and A2 were dominated 
by very young stellar populations, it would be surprising  that $\sim$5\,Myr 
old WR stars are present. \citet{Hunter2001} have 
presented Infrared Space Observatory (ISO) imaging and spectroscopy
of IC~4662, while \citet{Gilbert2008} discuss Spitzer imaging and
spectroscopy of these regions, supporting the claim of deeply embedded 
ionizing clusters by \citet{Johnson2003}. \citet{McQuinn2009} have 
analysed Hubble Space Telescope (HST) imaging of IC~4662 from which
they have estimated its recent star formation history.


In this paper we use Very Large Telescope (VLT) 
optical imaging and spectroscopy, supplemented by archival HST imaging to investigate the young, massive stellar content
of IC~4662. This paper is organised as follows: 
VLT optical imaging and spectroscopy 
of IC~4662, and archival HST imaging are presented in Section 2. A nebular 
analysis is presented in Section 3, followed by an analysis of the massive 
stellar content of IC~4662 in  Section 4. A discussion of the optical 
versus radio-derived  properties of the giant H\,{\sc ii} regions is 
presented in Section 5, together with a comparison of the star formation 
rate (SFR) of IC~4662 with other nearby metal-poor star forming galaxies.
Brief conclusions are reached in Section~6.


\begin{figure}
\begin{center}
\includegraphics[width=0.4\textwidth]{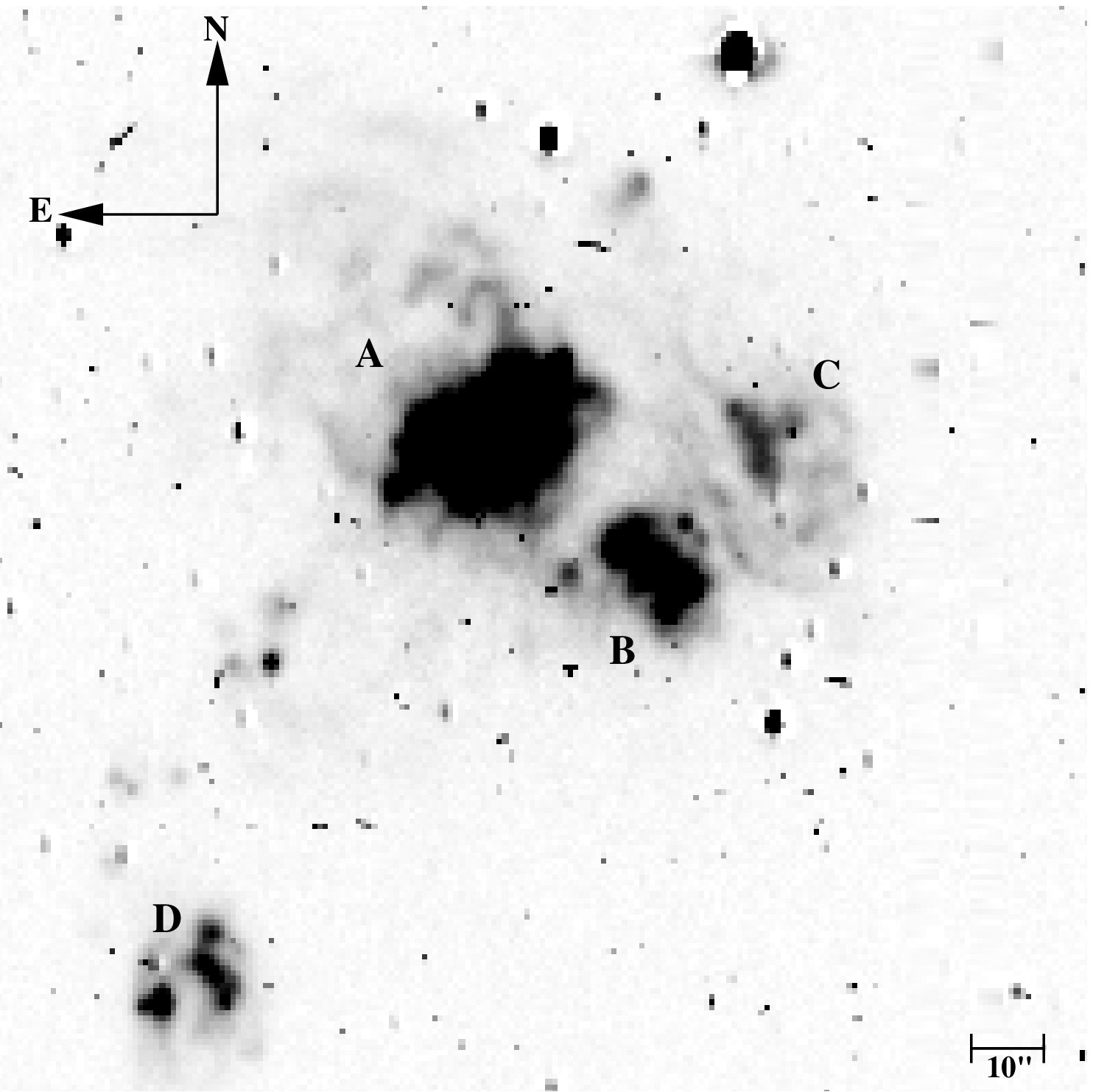}
\includegraphics[width=0.4\textwidth]{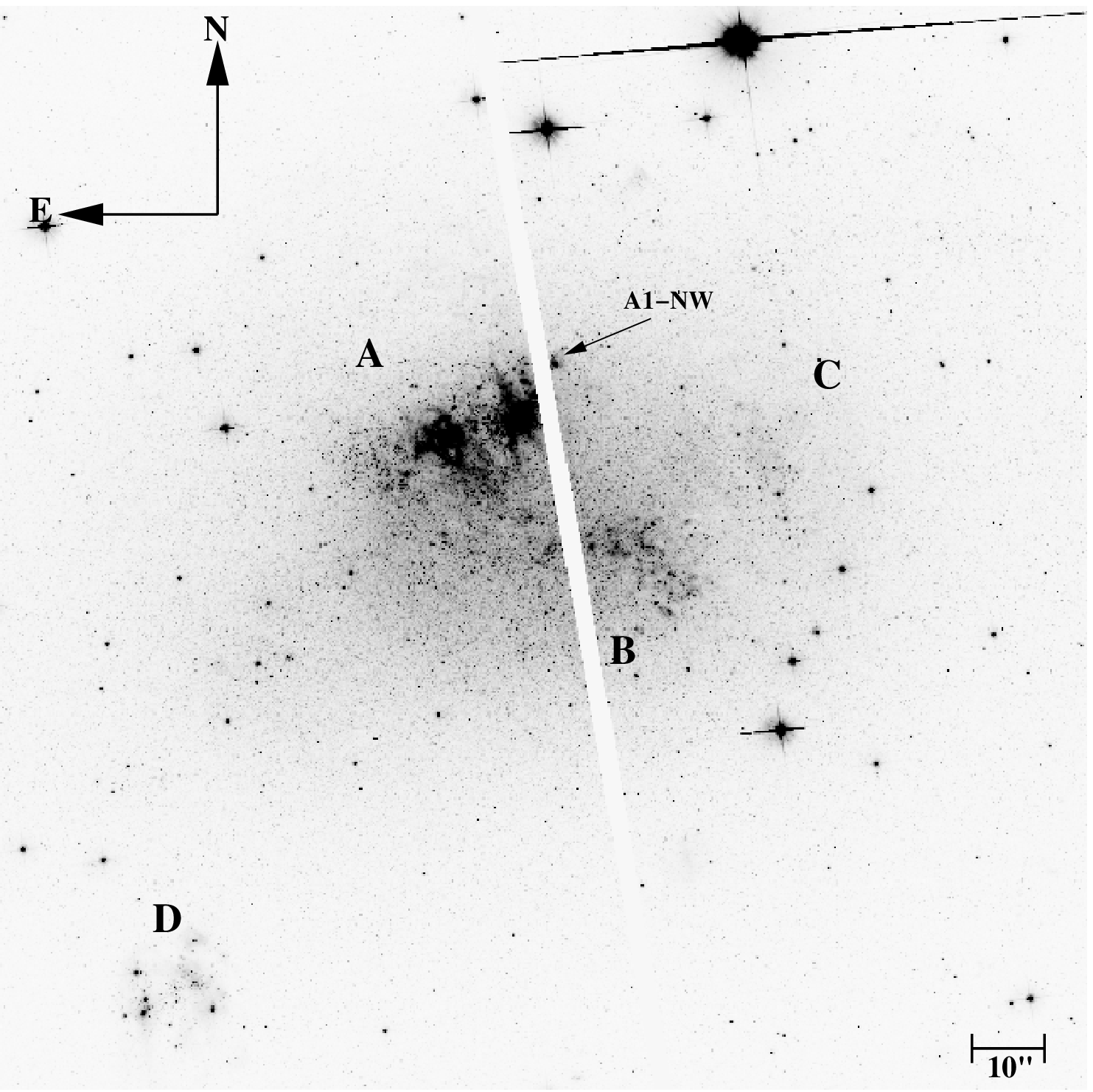}
\caption{2.5 $\times$ 2.5 arcmin (1.8 $\times$ 1.8 kpc at 2.44 Mpc) 
image of IC~4662 from CTIO 0.9m (H$\alpha$+[N\,{\sc ii}], upper, see 
\citet{Kennicutt2008}) and HST ACS/WFC (F606W, lower, see
\citet{Karachentsev2006}).}
\label{allregions}
\end{center}
\end{figure}

\section{Observations and Data Reduction}

Imaging and spectroscopy were obtained with the European Southern
Observatory (ESO) VLT in September 2000, using the 
Focal Reduced/Low dispersion Spectrograph $\#2$ (FORS2). These datasets 
were supplemented by
archival HST, Advanced Camera for Surveys (ACS) optical imaging.
We follow the nomenclature of \citet{Heydari1990} for the four
main H\,{\sc  ii}  regions of IC~4662, A, B, C and D throughout.

\subsection{VLT/FORS2 and CTIO 0.9m imaging}

In the absence of our own ground-based H$\alpha$ imaging of IC~4662, we have
been kindly provided with a continuum-subtracted H$\alpha$+[N\,{\sc ii}]
wide field image of IC~4662 obtained with the CTIO 0.9m telescope
on 14 Sep 2001 (J. Lee, priv. comm.) See \citet{Kennicutt2008} for 
further details of these observations. This image is presented in the 
upper panel of Figure~\ref{allregions}.  In addition to the
main giant H\,{\sc ii} region complex A, the figure highlights the fainter
H\,{\sc  ii} regions B and C to the south west and west of A,
plus region D, located 90~arcsec to the south east. The latter 
has previously been discussed by \citet{Hunter2001} (referred to as 
IC~4662-A, therein). The published H$\alpha$ flux of IC~4662 from 
\citet{Kennicutt2008} includes regions A--D, of which A1+A2 provide 
$\sim$70\% of the total H$\alpha$ emission.

\begin{figure*}
\begin{center}
\mbox{
    \subfigure{\includegraphics[width=0.4\textwidth]{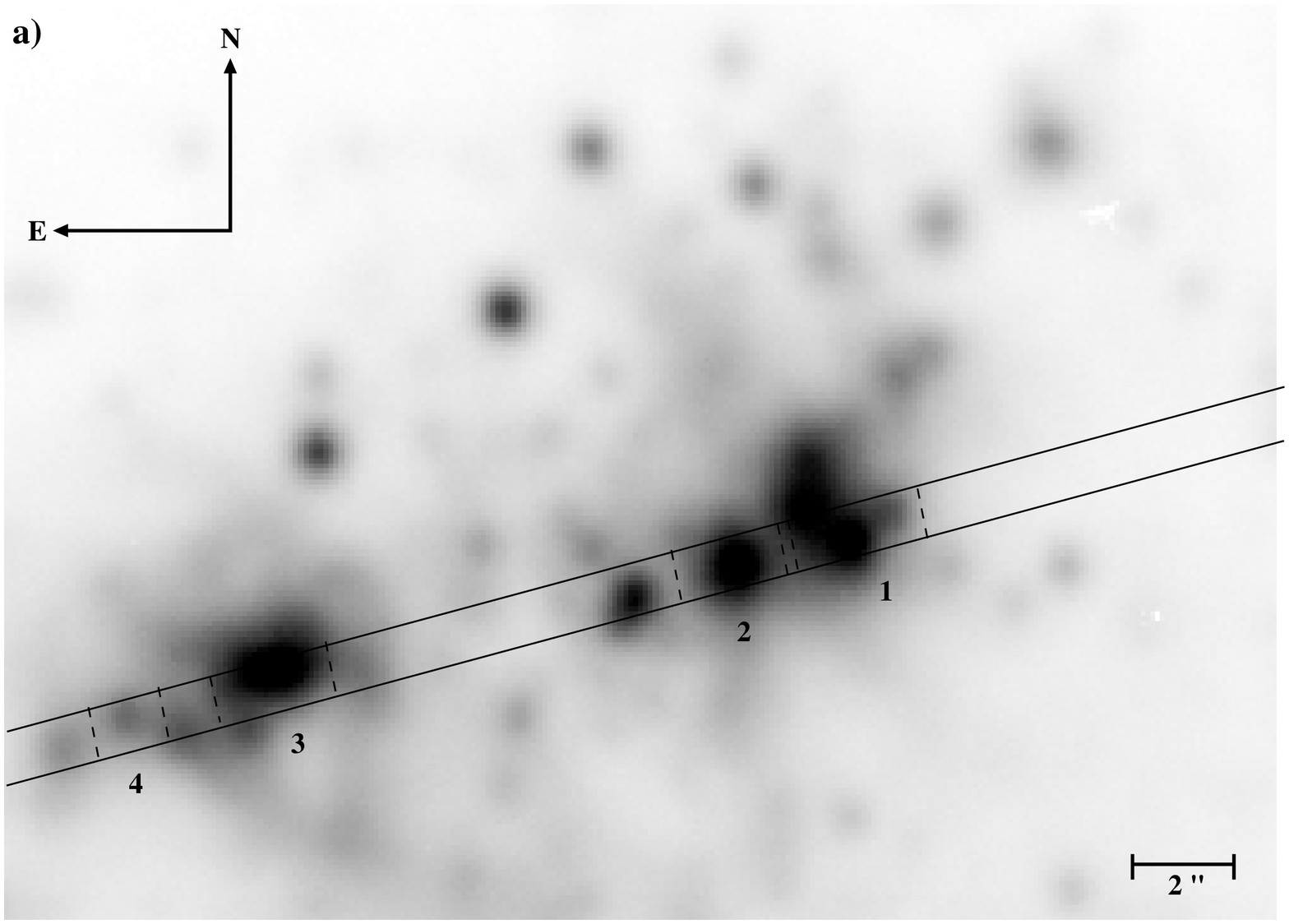}}
    \hspace*{0.2cm} 
    \subfigure{\includegraphics[width=0.4\textwidth]{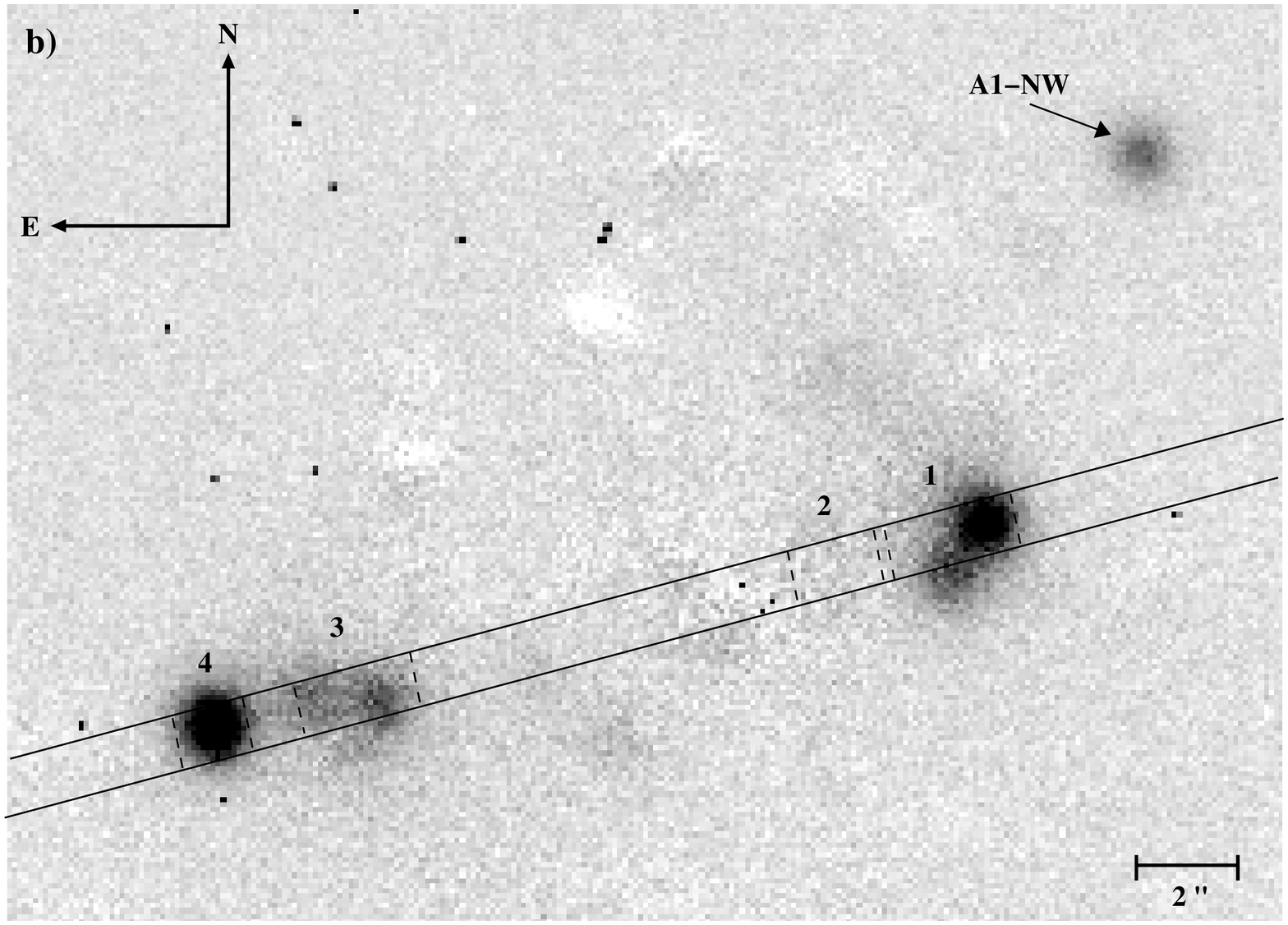}}
}
  \subfigure{\includegraphics[width=0.4\textwidth]{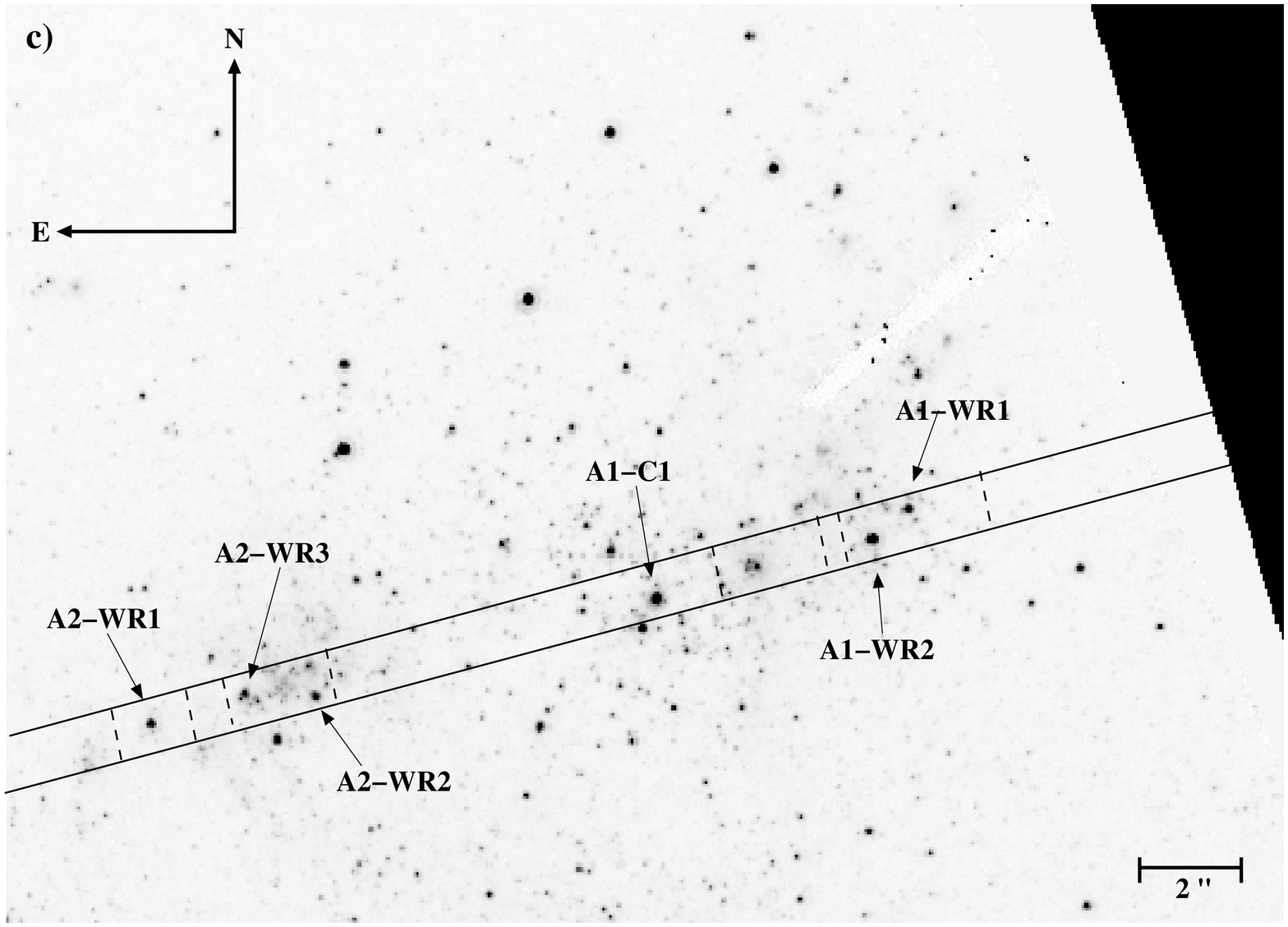}} 
  \hspace*{0.2cm}
  \subfigure{\includegraphics[width=0.4\textwidth]{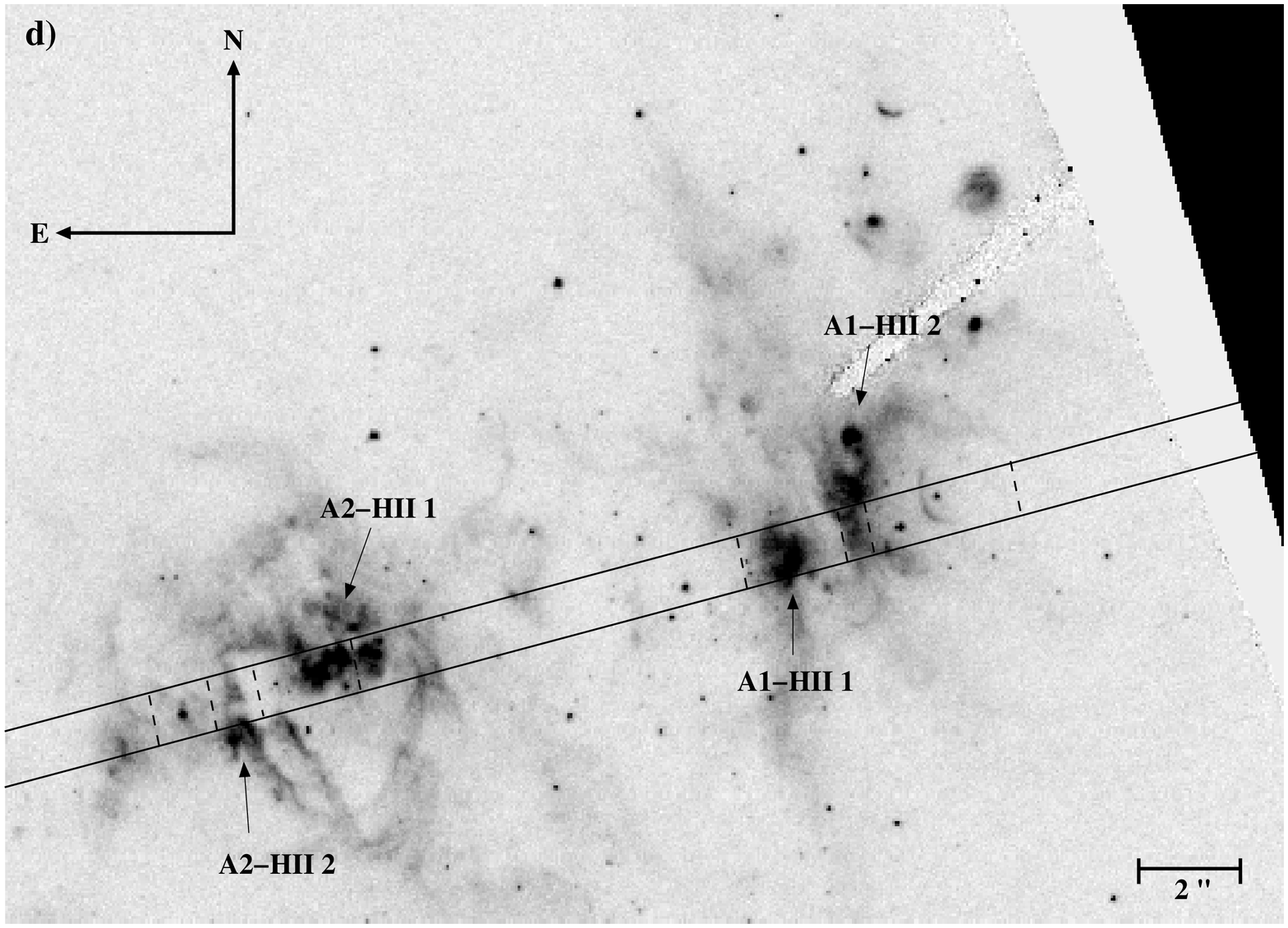}}
\caption{VLT/FORS2 Bessell V-band (panel a) and He\,{\sc ii} $\lambda$4686 
excess (panel b) images of the central $25\times18$ arcsec (300$\times$220 
pc at 2.44~Mpc) of IC~4662. The FORS2 long-slit has been overlaid, with 
extraction regions \#1 and 2 (within giant H\,{\sc ii} A1) and regions 
\#3 and 4 (within A2) marked.  Archival HST/ACS high resolution camera 
images of the same region   are presented in panel c (F550M) and d 
(F658N), whose footprint excluded  the North West region A1-NW. A number 
of continuum sources and H\,{\sc ii} regions are indicated (see 
Sect.~\ref{sect2.2}).}
\label{images}
\end{center}
\end{figure*}

A Bessell V-band ($\lambda_{c}$ = 554 nm, $\Delta
\lambda$ = 111 nm) image of IC~4662 was obtained with VLT/FORS2 on 4 Sep
2000 with an exposure time of 120~s, 
during moderate seeing conditions ($\sim$0.75 arcsec, 
airmass $\sim$1.4). The high resolution collimator provided
a field of view of 3.4 $\times$ 3.4 arcminutes (0.1
arcsec\,pix$^{-1}$). The detector was a single 2048 $\times$ 2048 Tektronix 
CCD with 24\,$\mu$m pixels.  A standard data reduction was applied,
involving bias subtraction, using \textsc{iraf} \citep{Tody1986}. The 
central $25 \times 18$ arcsec
region of IC~4662, corresponding to $300 \times 220$ pc, is presented in 
Fig~\ref{images}(a), showing multiple sources within each of the giant 
H\,{\sc ii} regions A1 and A2 from \citet{Heydari1990}.

In addition, 
interference filter imaging was obtained at the same time,  centred upon the 
He\,{\sc ii} $\lambda$4686 line 
($\lambda_{c}$ = 469.1 nm, $\Delta \lambda$ = 6.4 nm) and the
adjacent continuum ($\lambda_c$ = 478.9 nm, $\Delta \lambda$ = 6.6 nm).
Exposure times were 600~s in both cases.
The continuum subtracted He\,{\sc ii} image is presented in 
Fig.~\ref{images}(b), revealing two bright  $\lambda$4686 emission-line regions, one
each within A1 and A2, plus two faint emission-line regions. One
is $\sim$3 arcsec to the west of the bright source in A2, while the
other is offset $\sim$8 arcsec north west from the bright source in A1.
We  refer to  the latter source as A1-NW - previously the subject of a study by 
\citet{Richter1991} - since it also hosts a 1 arcsec  diameter H\,{\sc ii} 
region\footnote{A1-NW is not  spatially coincident with the compact radio 
continuum source  IC~4662-N from \citet{Johnson2003} which lies 4 arcsec
further to the east}.

\subsection{HST/ACS imaging}\label{sect2.2}

We have retrieved high-resolution archival HST/ACS imaging of IC~4662 to 
supplement our ground-based datasets. Two separate programmes have 
obtained HST/ACS imaging, namely GO programme 10609 (PI W. D. Vacca) for 
which
High Resolution Camera (HRC) imaging (0.03 arcsec\,pix$^{-1}$) were 
obtained
using the F330W, 
F435W, F550M, F658N and F814W filters with exposure times between 
400--2000 sec. In addition, ACS Wide Field Camera (WFC) imaging (0.1 
arcsec\,pix$^{-1}$) has been obtained under GO programme 9771 (PI I. D. 
Karachentsev) using the F606W (1200 sec) and F814W (900 sec) filters 
\citep{Karachentsev2006}. 

In the lower panel of Figure~\ref{allregions}, 
we present the ACS/WFC F606W image of 
IC~4662 on the same scale as the ground-based CTIO H$\alpha$+[N\,{\sc ii}] 
image, which
reveals that stellar populations in region D are physically separated from the 
remainder of the galaxy. Coordinates of individual sources were
obtained from this ACS/WFC image, using a fit to 7 nearby GSC-2 reference 
stars in the \textsc{starlink} image analysis package \textsc{gaia}. 
\citet{Johnson2003}  have previously presented  archival HST/WFPC2 F300W imaging of the 
central region of IC~4662.

In Fig.~\ref{images}(c) we present the ACS/HRC F550M image of
the central region of IC~4662, sampling the same region as that presented
in Fig.~\ref{images}(a) from VLT/FORS2. The improvement in spatial 
resolution allows us to identify individual He\,{\sc ii} $\lambda$4686 
excess regions from the FORS2 datasets, two within A1 that we shall 
hereafter refer to as
A1-WR1 and A1-WR2 -- the latter 0.4 arcsec south of a much brighter source --
plus three regions within A2, namely A2-WR1, A2-WR2 and 
A2-WR3. Sources A1-WR1, A1-WR2 and A2-WR1 are point-like, whereas sources 
A2-WR2 and A2-WR3  are spatially extended.

It is apparent from Fig~\ref{images}(c) that the He\,{\sc ii} 
$\lambda$4686 emission region, A1-NW, to the north west does not lie within 
the footprint of the HRC datasets. Fortunately, it is included within the 
WFC imaging (Fig.~\ref{allregions}, lower panel), revealing two point-like sources 
$\sim$0.2 arcsec ($\sim$2.5 pc) apart, located at the centre of extended 
H$\alpha$ emission. Either (or both) of these may contribute 
to the $\lambda$4686 emission. The visually brighter, south-west component 
is referred to as A1--WR3, with the fainter, north-east source labelled 
as A1--WR4.

In addition, we have labelled a bright, extended
continuum source to the east of region A1 as A1-C1. This was included within 
our VLT/FORS2 long-slit, and allows us to absolutely flux calibrate our 
spectroscopy (see sect.~\ref{spect}). 

Finally, we present the HRC/F658N image of the 
central region of IC~4662 in Fig.~\ref{images}(d). Two bright knots of 
H$\alpha$ emission can be seen within each of A1 and A2, that we shall 
refer to as A1-HII~1, A1-HII~2, A2-HII~1 and A2-HII~2, respectively.
3\,cm peak intensities from \citet{Johnson2003} correspond to 
A1-HII~2 and A2-HII~1. Their continuum source IC4662-N corresponds to a faint 
arc located 6 arcsec ($\sim$75 pc) to the north of A1-HII~2 in 
Fig.~\ref{images}(d).


\begin{table*} \caption{Observed line fluxes, $F_{\lambda}$, and 
intensities, $I_{\lambda}$, of nebular and stellar (Wolf-Rayet, WR) lines 
in each of the four VLT/FORS2 IC~4662 apertures, relative to H$\beta$=100. 
H$\beta$ fluxes are shown separate from normalised values in units of 
10$^{-13}$ erg s$^{-1}$ cm$^{-2}$. } 
\begin{tabular}{
l@{\hspace{1.5mm}}c@{\hspace{1.5mm}}c@{\hspace{3mm}}
r@{\hspace{1.5mm}}r@{\hspace{4mm}}
r@{\hspace{1.5mm}}r@{\hspace{4mm}}
r@{\hspace{1.5mm}}r@{\hspace{4mm}}
r@{\hspace{1.5mm}}r} 
\noalign{\hrule\smallskip\hrule\smallskip}
$\lambda$($\AA$) & ID & & 
\multicolumn{2}{c}{\#1} & 
\multicolumn{2}{c}{\#2} & \multicolumn{2}{c}{\#3} & 
\multicolumn{2}{c}{\#4}\\

& & & $F_{\lambda}$ & $I_{\lambda}$ & $F_{\lambda}$ & $I_{\lambda}$ & 
$F_{\lambda}$ & $I_{\lambda}$ & $F_{\lambda}$ & $I_{\lambda}$ \\
\noalign{\smallskip\hrule\smallskip}
3727 & [O\,{\sc ii}]        &    & 66.2  & 83.8 & 102.2 & 113.9 & 146.7 & 174.8 & 142.2 & 163.6 \\ 
3867 & [Ne\,{\sc iii}]      &    & 40.6  & 50.1 & 57.6   & 62.6 &  10.1 & 11.8  & 183.6 &  204.7\\
3968 & [Ne\,{\sc iii}] + H7 &    & 24.8  & 30.1 & 22.4   & 24.2 &  6.2  &  7.1  & 115.0 & 127.1 \\
4058 & N\,{\sc iv}          & WR & 1.8   & 2.1   &   --  &  --  &  1.8  & 2.1   &  0.5  & 0.6   \\
4100 & H$\delta$            &    & 21.1  & 25.4 & 23.9  & 27.1  & 22.9  & 26.4  & 25.0  & 28.5 \\
4340 & H$\gamma$            &    & 41.4  & 46.4 & 45.9  & 49.9  & 42.3  & 45.5  & 42.8  & 46.4 \\
4363 & [O\,{\sc iii}]       &    & 7.1   & 7.9  & 6.4   & 7.4   & 6.4   & 7.0   & 6.8   & 7.4  \\ 
4659 & [Fe\,{\sc iii}]      &    & 0.41  & 0.43 & 0.47  & 0.49  & 6.3   & 6.5   & 0.57  & 0.59  \\
4686$^{\ast}$ & He\,{\sc ii}         & WR & 8.5   & 8.9  & 2.1   & 2.2   & 
4.17  & 4.25  & 14.6  & 15.3 \\
4712 & [Ar\,{\sc iv}]       &    & 2.0   & 2.1  & 1.5   & 1.5   & 0.96  & 0.99  & 1.2   & 1.2  \\
4741 & [Ar\,{\sc iv}]       &    & 0.96  & 0.99 & 0.70  & 0.72  & 0.36  & 0.37  & 0.54  & 0.56  \\
4861 & H$\beta$             &    & 100   & 100  & 100   & 100   & 100   & 100   & 100   & 100  \\ 
4959 & [O\,{\sc iii}]       &    & 213.8 & 206.3& 198.4 & 198.5 & 184.8 & 180.4 & 203.6 & 202.0 \\ 
5007 & [O\,{\sc iii}]       &    & 630.0 & 604.0& 583.0 & 590.6 & 542.9 & 528.6 & 604.4 & 607.8 \\
5411 & He\,{\sc ii}         & WR & 0.56  & 0.50 &  --   & --    &  --   &   --  &  1.16 & 1.15  \\
5808 & C\,{\sc iv}          & WR &   --  &  --  & --   &  --   &  2.7   &  2.4  &   --  & --    \\
 5876 & He\,{\sc i}          &    & 12.1  & 9.8  & 11.7 & 11.3  & 13.9   & 12.0  & 11.9  & 11.1 \\
\noalign{\smallskip\hrule\smallskip}
H$\beta$                   &     &       & 0.44 & 1.12 & 0.39 & 0.49 & 0.44 & 0.86 & 0.13 & 0.20 \\
\noalign{\smallskip\hrule\smallskip}
\multicolumn{3}{l}{$E(B-V)$} & 
\multicolumn{2}{c}{0.28$\pm$0.05} & \multicolumn{2}{c}{0.08$\pm$0.05} & 
\multicolumn{2}{c}{0.20$\pm$0.05} & \multicolumn{2}{c}{0.12$\pm$0.05} \\
\multicolumn{3}{l}{T$_{e}$ (K)} &
\multicolumn{2}{c}{12900$\pm$700} & \multicolumn{2}{c}{12700$\pm$900} & 
\multicolumn{2}{c}{12900$\pm$700}  & \multicolumn{2}{c}{12500$\pm$400} \\
\multicolumn{3}{l}{O$^{+}$/H ($\times$10$^{-5}$)} &
\multicolumn{2}{c}{1.20$\pm$0.07}  & \multicolumn{2}{c}{1.7$\pm$0.1}   
&\multicolumn{2}{c}{2.5$\pm$0.18}   & \multicolumn{2}{c}{2.6$\pm$0.2}   \\
\multicolumn{3}{l}{O$^{2+}$/H ($\times$10$^{-5}$)} &
\multicolumn{2}{c}{10.3$\pm$0.53}  
& \multicolumn{2}{c}{10.6$\pm$0.61}  & 
\multicolumn{2}{c}{9.0$\pm$0.65}   & \multicolumn{2}{c}{11.2$\pm$0.87}  \\
\multicolumn{3}{l}{12 + $\log$(O/H)}  & \multicolumn{2}{c}{8.06$\pm$0.02}  
& \multicolumn{2}{c}{8.09$\pm$0.02}  &
\multicolumn{2}{c}{8.06$\pm$0.03}  & \multicolumn{2}{c}{8.14$\pm$0.03}  \\
\noalign{\smallskip\hrule\smallskip}
\multicolumn{10}{l}{$\ast$: He\,{\sc ii}  $\lambda$4686 fluxes are the sum 
of nebular and stellar components.}
\end{tabular}
\label{fluxes}
\end{table*}

\subsection{VLT/FORS2 spectroscopy}\label{spect}

VLT/FORS2 long slit spectroscopy was also carried out on 4 Sep 2000,
using the high resolution collimator and 600B grism, centred at 465\,nm. 
From the He\,{\sc ii} $\lambda$4686 excess image (Fig.~\ref{images}b) we 
chose a position angle of 104 degrees east of north, 
to sample the two bright sources within
A1 and A2, which also permitted the fainter source within A2 to be observed,
though not A1-NW. 
A single 1800~s exposure was obtained using a 1 arcsec
slit, albeit during observing conditions that had deteriorated 
considerably both in airmass ($\sim 2$) and seeing ($\sim$2 arcsec).

In total, four apertures were extracted, as shown in 
Figure~\ref{images}. Aperture 1 was centred upon the bright 
$\lambda$4686 source in A1 (A1-WR1 plus A1-WR2 to its south-east), 
aperture 2 included the bright H\,{\sc ii} region (A1-HII~1), 
aperture 3 included the fainter $\lambda$4686 source 
in A2 (A2-WR2 and A2-WR3 to its east) and finally aperture 4 included
the bright $\lambda$4686 source in A2 (A2-WR1). Due to the poor quality of 
the observing conditions, we also extracted the region centred upon the 
continuum source A1-C1, in order to facilitate absolute flux calibration.

A standard data reduction followed, involving bias subtraction and
extraction of apertures using \textsc{iraf}. Wavelength calibration
was achieved using a reference He/Hg/Cd arc lamp, revealing a spectral
range of 3710--6170$\AA$, plus a spectral resolution of 7$\AA$ as
measured from arc lines. Unfortunately, the primary spectrophotometric
standard star was observed using a different central wavelength to the
target, so a secondary star, HD~5980, observed during the same night
with the same setup as IC~4662 was used in its place. An unpublished,
flux calibrated dataset of HD~5980 obtained with the Mt Stromlo 2.3m 
Dual Beam Spectrograph (DBS) in Dec 1997 enabled a relative flux 
calibration.

Absolute flux calibration for each aperture was achieved in two steps, 
given the non-photometric conditions of the spectroscopic datasets. 
First, the continuum flux of the
aperture centred upon A1-C1 was set to the average of its F435W and
F550M fluxes, which were obtained from photometry using 
\textsc{daophot}, the point-spread function (PSF) fitting routine within 
\textsc{iraf}. The resulting flux calibrated spectrum agreed with
the photometry to within 10\%, but of course relies upon A1-C1 dominating
the VLT/FORS2 aperture. Second, slit losses of the other apertures
(with respect to A1-C1) were estimated from the two dimensional spectral
image, indicating throughputs of 60--85\% parallel to the slit, and 
adopting a uniform 50\% transmission perpendicular to the slit.  Overall, 
agreement between the flux calibrated spectra and the flux inferred from 
bright sources within each aperture was satisfactory, suggesting errors of 
$\sim$25\%.

 \begin{figure}
   \centering
   \includegraphics[bb=30 100 475 725,width=0.7\columnwidth,angle=-90]{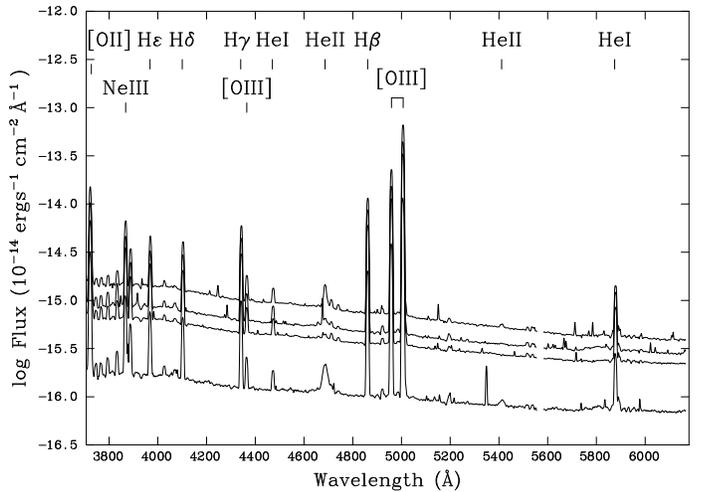}
     \caption{Flux calibrated, dereddened VLT/FORS2 spectra of 
IC~4662 apertures \#1, 3, 2 and 4 (from top to bottom).}
                   \label{figure2}
   \end{figure}

\section{Nebular Analysis}

In this section we shall derive the nebular properties of IC~4662 A1 and
A2 from VLT/FORS2 spectroscopy and HST/ACS F658N imaging, and present
comparisons with previous results.

\subsection{Properties of A1 and A2 from spectroscopy}

Our analysis was performed using gaussian fits to the observed high 
excitation nebular (and stellar) emission lines using the Emission Line 
Fitting (\textsc{elf}) routine within the 
\textsc{starlink} package 
\textsc{dipso}. These are presented in Table \ref{fluxes}.

The interstellar extinction for each aperture was determined using the 
observed Balmer line ratios F(H$\gamma$)/F(H$\beta$) and 
F(H$\delta$)/F(H$\beta$) together with Case B recombination theory 
\citep{Hummer1987}, assuming n$_{e}\sim$100 cm$^{-3}$ and 
T$_{e}\sim$10$^{4}$K, plus a standard Galactic extinction law 
\citep{Seaton1979}. Extinctions ranged from 0.08 mag, just above the 
foreground extinction of $E(B-V)$=0.07 mag \citep{Schlegel1998}, to 0.28 
mag, with average values of $E(B-V)$ = 0.162 and 0.165 mag for regions A1 
and A2, respectively. We assign an uncertainty of $\pm$0.05 to these 
values, given problems associated with flux calibration and the neglect of 
underlying (stellar) absorption lines. Consequently, these are consistent 
- within the quoted errors - with values of 0.11 and 0.14 mag for A1 and 
A2 from \citet{Hidalgo2001}. \citet{McQuinn2009} estimated an extinction
of $A_{\rm V} \approx 3.1 E(B-V) \approx$ 0.3 -- 0.4 towards IC~4662 from
their analysis of HST/ACS imaging.

We present the de-reddened spectra of apertures \#1--4 in 
Fig.~\ref{figure2}. In spite of the low spectral resolution, strong 
[O\,{\sc iii}] $\lambda$4363 emission is observed in all cases, allowing a 
determination of the nebular temperature. Emission line fluxes and 
intensities are presented in Table~\ref{fluxes}, including both strong 
nebular and stellar lines, the latter indicated with `WR', 
although He\,{\sc ii} $\lambda$4686 may be of stellar, or nebular, origin 
(see Sect.~\ref{sect4.2}).


Using the [O\,{\sc iii}](4959+5007)/4363$\AA$ line intensity ratio 
\citep{Osterbrock2006} we calculate T$_e$ for each region, for which we 
find T$_{e}\sim$12700 K in each case (Table \ref{fluxes}), similar to 
T$_{e}\sim$12000 K from \citet{Hidalgo2001}. Unfortunately, our VLT/FORS2 
spectroscopy is of insufficient resolution to separate the [O\,{\sc ii}] 
$\lambda$3726, 3729 doublet and does not cover the [S\,{\sc ii}] 
$\lambda$6717, 6731 doublet, so we adopt a typical H\,{\sc ii} region 
electron density of $\sim$100 cm$^{-2}$. For comparison, \citet{Gilbert2008} 
obtained $n_{e}$ = 150 (212) cm$^{-2}$ for A1 (A2) from the observed
[S\,{\sc iii}] 18.7$\mu$m/33.5$\mu$m ratio in their Spitzer/IRS datasets.
Emissivities were then determined  for each aperture using \textsc{fivel} 
\citep{deRobertis1987}, from which 
the oxygen abundance was calculated in each case. We find an average 
metallicity of 12 + $\log$(O/H) $\sim$ 8.1, close to that of the SMC 
\citep{Russell1990}, and in excellent agreement with the previous 
determinations of A1 and A2 by 
\citet{Heydari1990} and \citet{Hidalgo2001}. Oxygen 
abundances for individual apertures are also presented in 
Table~\ref{fluxes}. These are consistent with
metal contents of 0.12--0.17 $\times$ solar for neon and sulphur 
obtained by \citet{Gilbert2008} from Spitzer/IRS spectroscopy of A1 and A2.
Note that \citet{Hidalgo2001} obtain a substantially
lower  oxygen abundance of $\log$(O/H)+12 = $7.7$ for region D.

\subsection{Properties of A1 and A2 from imaging}

The ionizing fluxes of giant H\,{\sc ii} regions IC~4662-A1 and A2 
may be estimated from their H$\alpha$
luminosity \citep{Kennicutt2008}. Aperture photometry of the ACS/HRC
F658N image (recall Fig.~\ref{images}d) allows the H$\alpha$+[N\,{\sc ii}] 
fluxes of A1 and A2 to be measured, using radii of 6.15 arcsec (75 pc)
and 5.5 arcsec (65~pc), respectively, under  the reasonable  assumption 
that the stellar contribution to this filter is 
negligible. We defer a discussion of the global star formation rate of 
IC~4662 until 
Section~5.

Table~\ref{A1/A2} presents observed H$\alpha$ fluxes for these
regions, corrected for both background and aperture size 
\citep{Sirianni2005}. 
We also account for the contribution of [N\,{\sc ii}] to the 
F658N flux using the  ratio F[N\,{\sc ii}]/F(H$\alpha) \sim$ 0.03 obtained 
by \citet{Hidalgo2001}. For our adopted distance of 2.44 Mpc, the
\citet{Kennicutt1998} conversion from H$\alpha$ luminosity to ionizing 
flux, $Q_0$, implies  145$\times$10$^{49}$s$^{-1}$ and 
121$\times$10$^{49}$s$^{-1}$ for A1 and A2, respectively. This equates to 
145 and 121 equivalent O7V stars, based upon an ionizing flux of $\sim$ 
10$^{49}$ photons s$^{-1}$ for SMC-metallicity O7V stars \citep{Hadfield2006}. 

In Table~\ref{A1/A2} we compare these values to those obtained from 3 
cm/8.6 GHz continuum fluxes \citep{Johnson2003}, and Hu~$\alpha$ 
(12.4$\mu$m) line fluxes from Spitzer/IRS spectroscopy 
\citep{Gilbert2008}. The optically derived ionizing flux of A1 is 
approximately 2/3 of the radio derived value, while that of A2 is 
approximately 1/2 of the 3 cm determination, supporting previous 
suggestions of deeply embedded sources within A1 and A2, 
albeit contributing at most 30-50\% of the total ionizing flux.

In A1, the Spitzer/IRS derived Lyman continuum flux is apparently higher
than that from the radio continuum. However, the Hu~$\alpha$ line appears unusually
broad in the IRS dataset with respect to other nebular lines, suggesting
a blend with another feature, which could account for up to 50\% of the observed 
Hu~$\alpha$ flux, bringing it closer into line with value obtained from 
H$\alpha$ (W.D. Vacca, priv. comm.).

\section{Analysis of individual sources}

In this section we use our VLT/FORS2 spectroscopy to estimate the 
massive star content of each region, supplemented by narrow-band
imaging datasets for region A1-NW which was not observed spectroscopically.
We also infer properties of individual ACS/HRC and WFC sources, for
which slit spectroscopy is available.

\subsection{O star content}

In the previous section we have obtained estimates of the equivalent number
of O7V stars for A1 and A2 from ACS/HRC H$\alpha$ imaging. In reality, 
the O star content of individual regions of IC~4662 depends
upon the age via 
\begin{equation} 
N(O)=\frac{N_\mathrm{O7V}}{\eta_{0}}
\end{equation}
where $\eta_{0}$ is a sensitive function of age \citep{Vacca1994, 
Schaerer1998}. 
Ages can be estimated from the equivalent width of Balmer lines, such as
H$\beta$ in our VLT/FORS2 spectra.
Table~\ref{content} presents the measured W$_{\lambda}$(H$\beta$) for each 
of our four apertures. Combining individual apertures reveals
W$_{\lambda}$(H$\beta$)$\sim$150$\AA$ and 190$\AA$ for A1 and A2,
 respectively. Slit spectroscopy of these regions by \citet{Heydari1990} revealed lower
equivalent widths of 146 and 122$\AA$, using a 2 arcsec slit width.

From SMC-metallicity evolutionary synthesis models \citep{Schaerer1998}, 
such line strengths suggest characteristic ages of 4.7 and 4.5 Myr for the 
regions sampled within A1 and A2, respectively, neglecting the minor 
contribution of weak stellar H$\beta$ absorption. In reality, young 
clusters will only contribute a fraction of the total continuum light 
sampled by the ground-based VLT/FORS2 spectroscopy, so the actual ages of 
clusters  are likely to be somewhat younger than that inferred. 
Nevertheless,  $\eta_{0}\sim$  0.2--0.25 follows from the measured H$\beta$ 
equivalent  widths, allowing estimates of the number of O stars, N(O), as 
presented in Table~\ref{content}.

These estimates serve as upper limits since the contribution of Wolf-Rayet
stars to ionizing fluxes are neglected, and are solely valid for the 
sub-regions of A1 and A2 that are sampled by the individual apertures 
(recall Fig~\ref{images}a). Therefore 
one should not infer global O star numbers for A1 and A2 from our age
estimates, since there will doubtless be a spread in ages of sources
within these regions. For A1--C1, a significantly  
smaller H$\beta$ equivalent width of $\sim$48 $\AA$\ was  measured, 
suggesting a greater age of $\sim$6 Myr.


\begin{table}
\caption{Observed H$\alpha$ fluxes of IC4662~A1 and A2 from optical HST/F658N 
imaging (6 arcsec radii), compared to radio and mid-IR results.}
\begin{tabular}{l@{\hspace{1.5mm}}c@{\hspace{3mm}}c@{\hspace{3mm}}c}
\noalign{\hrule\smallskip\hrule\smallskip}
Quantity          & Units                          &       A1              & A2                  \\
\noalign{\smallskip\hrule\smallskip}
$F$(H$\alpha$+[N\,{\sc ii}]) &(erg s$^{-1}$ cm$^{-2}$)   & 
1.94$\times 10^{-12}$ & 1.61 $\times 10^{-12}$ \\
$F$[N\,{\sc ii}]/$F$(H$\alpha$) & & 0.03                   & 0.03             
\\
$F$(H$\alpha$)&(erg s$^{-1}$ cm$^{-2}$)         & 1.89$\times 10^{-12}$ & 
1.56 $\times 10^{-12}$  \\
$E(B-V)$     & (mag)                               &  0.162                
& 
0.165   \\
$I$(H$\alpha$) & (erg s$^{-1}$ cm$^{-2}$)         & 2.75$\times$10$^{-12}$ 
& 
2.30$\times$10$^{-12}$   \\ 
$L$(H$\alpha$) & (erg s$^{-1}$)                  & 1.96$\times$10$^{39}$  
& 
1.63$\times$10$^{39}$    \\ 
$Q_0$(H$\alpha$) & (10$^{49}$ s$^{-1}$)         &145                    & 
121\\
\noalign{\smallskip\hrule\smallskip}
$Q_0$(3\,cm)$^{\ast}$ & (10$^{49}$ s$^{-1}$)             & 227                   
& 
242
\\
$Q_0$(Hu$\alpha$)$\ddag$ & (10$^{49}$ s$^{-1}$)     & 345                   
& 256 
\\
\noalign{\smallskip\hrule\smallskip}
\multicolumn{4}{l}{$\ast$: \citet{Johnson2003} adjusted to 2.44 Mpc.}\\
\multicolumn{4}{l}{$\ddag$: \citet{Gilbert2008} adjusted to 2.44 Mpc.}
\end{tabular}
\label{A1/A2}
\end{table}
 
\subsection{Wolf-Rayet populations}\label{sect4.2}

In Fig.~\ref{figure3} we present spectroscopy of the four VLT/FORS2
apertures around the He\,{\sc ii} $\lambda$4686 line. From comparison
with adjacent [Fe\,{\sc iii}] and [Ar\,{\sc iv}] nebular lines, the 
observed He\,{\sc ii} lines comprise (narrow) nebular and (broad)
stellar components, the latter from Wolf-Rayet stars.

We can estimate the Wolf-Rayet populations within individual apertures 
from the stellar He\,{\sc ii} $\lambda$4686 line luminosity, so we have 
taken care to distinguish between nebular and stellar components. Stellar 
line luminosities for regions \#1, \#3 and \#4, lie in the range 
1.8--4.4$\times10^{36}$ erg\,s$^{-1}$, with FWHM(He\,{\sc 
ii})$\approx$20--30$\AA$.  N\,{\sc iv} $\lambda$4058 emission is
observed in these regions, with N\,{\sc iii} $\lambda$4634-41 weak or 
absent, suggesting mid-type nitrogen sequence (WN5--6) spectral types. 

\citet{Crowther&Hadfield2006} present line luminosities for individual LMC 
and SMC WN5--6 stars. From a sample of 15 LMC WN5--6(ha) stars, the mean 
He\,{\sc ii} line luminosity is 1.75$\times 10^{36}$ erg\,s$^{-1}$. 
Therefore, the observed line luminosities of \#1, \#3 and \#4 are each 
consistent with 1--2 luminous, hydrogen-rich WN5--6ha stars 
(Table~\ref{content}). Alternatively, we also consider the possibility 
that sources are clusters hosting {\it multiple} hydrogen-poor WN stars. 
In view  of the measured oxygen abundance for IC~4662, we would instead
adopt the  4.3$\times 10^{35}$ erg\,s$^{-1}$ mean of 2 SMC WN5--6 stars, 
from which  4--10 (classical) WN5--6 stars are inferred, omitting any 
contributions from WC stars.

Weak, broad C\,{\sc iv} $\lambda$5801--12 from early-type WC stars is 
observed solely in region \#3. Unfortunately, no WC4 stars are observed in 
the SMC, so we face the choice of adopting a mean C\,{\sc iv} line 
luminosity of 3.3$\times 10^{36}$ erg\,s$^{-1}$ from LMC WC4 stars,
or that of the sole SMC WO star. O\,{\sc vi} $\lambda$3811--34 does not appear to be 
present in our VLT/FORS2 spectroscopy, favouring a single WC4 star in 
region 3. The C\,{\sc iv} FWHM is  difficult to  measure, since it is 
blended  with weak [N\,{\sc ii}]  $\lambda$5755  emission, but supports a 
WC4 subtype since FWHM$\sim$80 $\AA$. In addition, from the observed ratio 
of the C\,{\sc iii} $\lambda$4650 + He\,{\sc  ii}  $\lambda$4686  blend 
to the C\,{\sc iv} $\lambda$5801--12 
line luminosity in LMC WC4 stars,  one would expect a contribution of 
$\sim 2 \times 10^{36}$  erg\,s$^{-1}$ to the $\lambda$4650/4686 line 
luminosity in region 3 from the WC4 star. This is similar to that  observed,
and  would suggest that WN stars do not contribute to this region. However,
the observed peak wavelength of the blue feature in region \#3 lies at 
$\lambda$4686 rather than $\lambda$4650, arguing against  this 
interpretation. In fact, two separate sources are observed within 
region \#3, namely A2-WR2 and A2-WR3. One may host a WC star and 
the other a WN-type star.


In addition to stellar He\,{\sc ii} $\lambda$4686, \#1--3 reveal
nebular He\,{\sc ii} emission, with I(He\,{\sc ii})/I($H\beta$) = 0.03
in region 1 and $\leq$0.01 in regions 2--3. \citet{Hidalgo2001} obtained
$I$(He\,{\sc ii})/$I$(H$\beta$) = 0.05 for stellar and nebular 
$\lambda$4686.  Other high ionization 
lines are observed in A1 and A2, namely [O\,{\sc iv}] 25.9$\mu$m 
from Spitzer/IRS \citep{Gilbert2008}. Since O stars are incapable of
producing significant numbers of high energy photons, either shocks
or Wolf-Rayet stars \citep{Hadfield2006} are likely to be responsible 
for the nebular He\,{\sc ii} and [O\,{\sc iv}] lines. The former could be
reconciled with the absence of WR stars in region 2 owing to 
spatially extended nebular He\,{\sc ii} emission from region 1.


Finally, although our VLT/FORS2 spectroscopy did not include the north
west region of A1, we have estimated its line flux through the
He\,{\sc ii} $\lambda$4686 filter from a comparison with the spectroscopic
line flux of region 4 (recall Fig.~\ref{images}b).  
We obtain a He\,{\sc ii} line flux of 10$^{-15}$ erg s$^{-1}$ 
cm$^{-2}$ for region A1-NW. \citet{Richter1991} have published 
spectroscopy of this source, revealing an early-type WC subtype. Neglecting 
any  potential nebular emission, for an assumed extinction of 
$E(B-V)$=0.07  mag we estimate a luminosity of $\approx 10^{36}$ 
erg\,s$^{-1}$ in the
469.1\,nm filter. For WC stars, approximately half of the
C\,{\sc iii} $\lambda$4650/He\,{\sc ii} $\lambda$4686 luminosity 
contributes to the 469.1\,nm filter, so the total 
blue luminosity is $\approx 2 \times 10^{36}$ erg\,s$^{-1}$, consistent 
with a sole WC star. We are unable to discriminate between the two 
potential sources (A1-WR3 
and A1-WR4) from our ground-based imaging since they are only 0.2 arcsec
($\sim$2pc) apart in ACS/WFC datasets and possess similar 
$m_{\rm F606W}$ -- $m_{\rm F814W}$ colours.


Our estimate of the stellar content of A1-NW is  significantly lower 
than previous ESO 2.2m 
narrow-band imaging and  3.6m/IDS spectroscopy of this region by 
\citet{Richter1991}. They obtained 
an integrated C\,{\sc iv} $\lambda$4650/He\,{\sc ii} $\lambda$4686 flux of 
3$\times$10$^{-14}$ erg s$^{-1}$ cm$^{-2}$, from which $\sim$10 late WN 
and  early WC stars were inferred. For a solely foreground extinction, 
their quoted nebular H$\beta$ flux would imply L(H$\beta$) = 1.1$\times 
10^{38}$ erg\,s$^{-1}$ or $Q_{0} = 2.3\times 10^{50}$ photon s$^{-1}$.
The reason for this substantial difference is unclear.

\begin{table}
\caption{VLT/FORS2 nebular and stellar properties of apertures 1--4 within
IC4662~A1 and A2, including estimated O star numbers, N(O), and 
luminosities of nebular and Wolf-Rayet lines.}
\begin{tabular}{l@{\hspace{0mm}}c@{\hspace{0mm}}c@{\hspace{1.5mm}}c
@{\hspace{1.5mm}}c@{\hspace{1.5mm}}c}
\noalign{\hrule\smallskip\hrule\smallskip}
&  & \multicolumn{2}{c}{A1} & \multicolumn{2}{c}{A2} \\

              & Units          & \#1                   & \#2                   & \#3                   & \#4                   
\\
\noalign{\smallskip\hrule\smallskip}

$L$(H$\beta$) & (erg s$^{-1}$)   & 8.0$\times$10$^{37}$  & 
3.5$\times$10$^{37}$  & 6.2$\times$10$^{37}$  & 1.4 $\times$10$^{37}$ \\
$\log Q_0$ & (photon s$^{-1}$) & 50.22                 & 49.87                 
& 50.11                 & 49.47                \\
W$_\lambda$(H$\beta$) & ($\AA$)& 138                   & 159                   & 177                   & 197                   \\
$\eta_{0}$            &        & 0.2                   & 0.25                  & 0.25                  & 0.25                  \\
N(O)                  &        & 84                    & 30                    & 52                    & 12                     \\

$L$(He\,{\sc ii}$^{\rm WR}$ 4686) & (erg\,s$^{-1}$) &  
4.4$\times$10$^{36}$  
&    --                  & 2.1$\times$10$^{36}$  & 1.8$\times$10$^{36}$ \\
$L$(He\,{\sc ii}$^{\rm neb}$ 4686) & (erg\,s$^{-1}$)& 2.7$\times$10$^{36}$   
& 5.3$\times$10$^{35}$   & 3.0$\times$10$^{35}$  & --  \\
$L$(C\,{\sc iv} 5801)  & (erg\,s$^{-1}$)            &   --                   
&           --           & $\leq$1.5$\times$10$^{36}$  & --     \\
\noalign{\smallskip\hrule\smallskip}
\end{tabular}
\label{content}
\end{table}

\begin{table*}
\caption{Vega magnitudes from ACS/HRC (ACS/WFC for A1-WR3 and 
A1-WR4) imaging, plus inferred  cluster properties from 
\textsc{starburst99} \citep{Leitherer1999} models, for
ages of $\sim$4.5 Myr ($\sim$6 Myr for A1-C1).}
\begin{tabular}{c@{\hspace{0mm}}c@{\hspace{-2mm}}c@{\hspace{2mm}}c
@{\hspace{2mm}}c@{\hspace{2mm}}c@{\hspace{2mm}}c@{\hspace{2mm}}c
@{\hspace{2mm}}c@{\hspace{2mm}}c}
\noalign{\hrule\smallskip\hrule\smallskip}
Source   & &                       A1--WR1 & A1--WR2   & A1--C1 & A2--WR1 & A2--WR2 & A2--WR3 & A1--WR3 & A1--WR4\\
Aperture & & \#1     & \#1     &        & \#4     & \#3   &\#3 & (A1-NW) & (A1-NW) \\
\noalign{\smallskip\hrule\smallskip}
RA (J2000)                       &&17:47:08.66&17:47:08.77&17:47:09.45&17:47:11.03&17:47:10.52&17:47:10.74&17:47:08.18&17:47:08.20\\
Dec (J2000)                      &&--64:38:17.1&--64:38:18.1&--64:38:18.8&--64:38:21.2&--64:38:20.6&--64:38:20.6&--64:38:09.8&--64:38:09.7\\
$m_{\rm F550M}$ &(mag)                & 19.53  & 20.82     & 18.72  & 19.25  & 19.85  & 19.97 & 20.17$^{1}$ & 20.29$^{1}$ \\
$m_{\rm F435W} - m_{\rm F550M}$ &(mag)& --0.18 & --0.35    & --0.18 & --0.18 & --0.09 & 0.16  & --          & --\\
$E(B-V)$  & (mag)                     & (0.07) & (0.07)    & (0.07) & (0.07) &  0.08  & 0.33  &  (0.07)     & (0.07) \\
M$_{\rm F550M}$ & (mag)               & --7.6  &  --6.3    & --8.4  & --7.9  & --7.3  & --7.9 & --7.0$^{1}$ & --6.9$^{1}$ \\
Mass & (10$^{3}$M$_\odot$)            &        &           & 3.4    &        & 1.3    & 2.3   &         \\
WR?    &                              & 1--2$\times$WN5-6ha & 
WN5--6?  & ---  &  1--2$\times$WN5-6ha  & WN5--6 or WC & WN5--6 or WC & 
WC? & WC? \\
\noalign{\smallskip\hrule\smallskip}
\multicolumn{10}{l}{Note: Adopted interstellar extinctions are shown in 
parenthesis.}\\
\multicolumn{10}{l}{$^{1} m_{\rm F606W}$}
\end{tabular}
\label{magnitude}
\end{table*}

\subsection{Clusters or single stars?}

Our VLT/FORS2 imaging allows us to identify the location of $\lambda$4686
emission line sources,  plus spectroscopy allows crude estimates of the 
massive stellar content of  individual apertures. However, 
recalling Figure~\ref{images}(c), ACS/HRC imaging  reveals multiple 
sources within these apertures.  Such sources are either compact 
($\sim$pc) clusters or individual stars.

We performed photometry  of ACS/HRC F435W  and F550M imaging with 
\textsc{daophot} to  obtain Vega magnitudes 
of individual sources, $m_{\rm F435W}$ and $m_{\rm F550M}$, following 
\citet{Sirianni2005}. In turn, absolute magnitudes follow from
our adopted distance of 2.44 Mpc and assumed intrinsic colours. 
In the case of sources within region A1-NW, which
fell beyond the footprint of the ACS/HRC datasets, aperture photometry
of radius 0.2 arcsec was performed using the ACS/WFC F606W image  
(Fig.~\ref{allregions}) using \textsc{gaia} 
 
Initially we assume uniform (foreground) interstellar extinctions of 
$E(B-V)$=0.07, from which we obtain absolute F550M (or F606W) magnitudes.
These lie in the range --6.3 to --8.4 mag for the WR sources in IC~4662.
This range spans that expected by individual stars (blue supergiants) 
and star clusters. We discriminate between these two cases from the 
ACS/HRC images. The majority of WR sources are point-like, with only 
A2-WR2,  A2-WR3 and A1-C1 spatially extended. Of course, it is  possible 
that  point sources could be compact clusters.  However, given the 
ACS/HRC spatial resolution of $\sim$0.05 arcsec (0.6 
pc at 2.44 Mpc), such clusters would need to be very compact to avoid 
detection.

The two  strong He\,{\sc ii} $\lambda$4686 emission line sources
are spatially coincident with point sources A1-WR1 ($M_{\rm F550M}$ = 
--7.6 mag) and A2-WR1 ($M_{\rm F550M}$ = --7.9 mag). Absolute 
visual magnitudes of individual WN stars typically lie in the range 
$M_{\rm B}$ = --4 to --7 mag. The only examples of visually bright WR
stars are the high mass, hydrogen-rich WN5--6ha stars for which $M_{\rm 
B}$  =  --7  to --8 mag \citep{CrowtherDessart98}. Therefore A1-WR1 and 
A2-WR1 are likely to be very luminous, single or binary WN5--6 stars. 
Their high He\,{\sc ii} $\lambda$4686 luminosities are consistent with 
1--2 luminous WN5--6 stars in each case. Such stars are located within 
compact, massive star clusters such as NGC~3603 in the Milky Way and R136 
in the LMC, but examples are also found within more modest star forming 
regions, such as WR20a  within Westerlund~2 in the Galaxy \citep{Rauw2004}.



A1-WR2, the other point source from aperture \#1 for which He\,{\sc ii} 
$\lambda$4686 emission  is detected, is rather fainter, with $M_{\rm 
F550M}$ =   --6.3 mag. Recalling Fig.~\ref{images}b, this source possesses 
a  He\,{\sc ii} line flux a factor of 2--3 times lower than A1-WR1, i.e.
it is likely to be a WN5--6 star that is more typical of the SMC field 
population \citep{Crowther&Hadfield2006}, once again either single  or 
within a  binary system, unless the emission from this source were solely 
nebular.

The absence of high spatial resolution continuum imaging of 
A1-NW at more than a single continuum band prevents us discriminating 
between A1-WR3 and A1-WR4 for the source of He\,{\sc ii} $\lambda$4686 emission
through their colours. Both are point sources at the resolution of 
ACS/WFC, of which A1-WR3 is the brightest with $M_{\rm F606W}$ = 
--7.0 mag. In view of spectroscopy of A1-NW by \citet{Richter1991}, which
revealed a characteristic WC4--5 spectral type, it is likely that one of
A1-WR3 or A1-WR4 is a binary system comprising a WC star and OB-type 
companion.

For the remaining -- spatially extended -- sources, A2-WR2, A2-WR3 (and 
A1-C1) we have estimated cluster properties by first obtaining revised 
absolute 
magnitudes through comparison with the colours expected for young massive 
clusters at ages of $\sim$4.5 Myr. This was obtained from nebular H$\beta$ 
line  equivalent widths for \textsc{starburst99} \citep{Leitherer1999}, 
based upon a  \citet{Kroupa2002} Initial  Mass Function (IMF, 0.1--100M$_{\sun}$) 
with a turnover at 0.5M$_{\sun}$,  and SMC metallicity Padova evolutionary 
models \citep{Fagotto1994}. Such  models are strictly valid when 
statistically large numbers of massive  stars are present within 
individual star clusters, and break down for low  cluster masses. 
Nevertheless, they provide a useful estimate of the likely  stellar masses 
of individual sources should they be compact star clusters. They agree 
closely with the statistical approach of \citet{Parker&Goodwin2007} 
and \citet{Furness2008} in  which stars are randomly sampled from the 
IMF.

Intrinsic $m_{\rm F435W}$ - $m_{\rm F550M}$ colours of $-$0.17
mag and $-$0.15 were, respectively, 
estimated for ages of 4.5 Myr (A2-WR2 and A2-WR3) and 
6 Myr (A1--C1) from {\textsc{starburst99}. Interstellar extinctions and 
absolute magnitudes are presented in Table \ref{magnitude}, together with
estimated cluster masses and the number of expected O and WR stars within
individual sources. We obtain cluster mass estimates no greater than a 
factor of two greater than the Orion Nebula Cluster (ONC) for which
$\sim 1800 M_{\odot}$ \citep{Hillenbrand1997}. One would expect 
between 3--5  O stars in such  clusters, and at most one WR star in any 
single cluster. A2-WR2 and A2-WR3 coincide with faint He\,{\sc ii} 
$\lambda$4686 emission in Fig.~\ref{images}(b) and so each likely hosts 
one SMC-type WN5--6 star or WC4 star.

 \begin{figure}
   \includegraphics[bb=25 85 505 725 ,width=0.7\columnwidth,angle=-90]{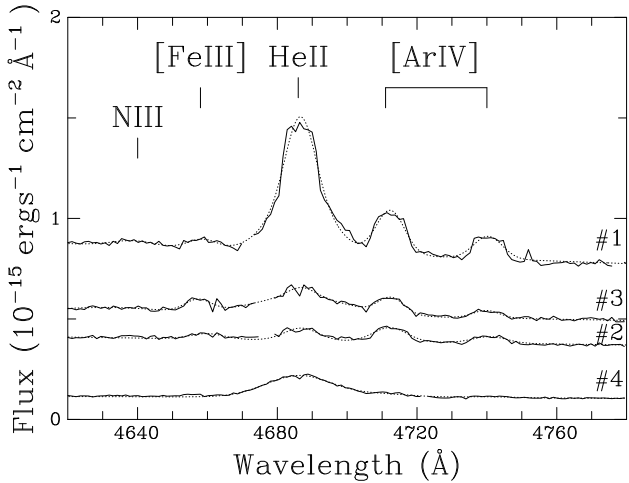}
     \caption{De-reddened VLT/FOR2 spectroscopy of apertures 1--4 in the
vicinity of He\,{\sc ii} $\lambda$4686 revealing both nebular and stellar 
components, plus nebular [Fe\,{\sc iii}] $\lambda$4658 and [Ar\,{\sc 
iv}] $\lambda$4711, 4741 (gaussian fits are shown as dotted lines).}
                   \label{figure3}
   \end{figure}

\begin{table*}
\caption{Comparison between IC~4662 H$\alpha$ fluxes from the 
literature. Inferred intensities, luminosities and  Lyman continuum 
fluxes, follow from an adopted distance of 2.44 Mpc, and include
3\,cm radio continuum derived ionizing flux from \citet{Johnson2003}. }
\begin{tabular}{l@{\hspace{1.5mm}}c@{\hspace{1.5mm}}c@{\hspace{1.5mm}}c@{\hspace{1.5mm}}c@{\hspace{1.5mm}}c@{\hspace{1.5mm}}c@{\hspace{1.5mm}}c@{\hspace{1.5mm}}r}
\noalign{\hrule\smallskip\hrule\smallskip}
                       & $\theta^{\ast}$ &
$F$(H$\alpha$+[N\,{\sc ii}])     & $F$[N\,{\sc ii}]/$F$(H$\alpha$) &  
$E(B-V)$  & 
$I$(H$\alpha$) &  $\log L$(H$\alpha$)    & $Q_0$  \\
  & (arcmin) & (erg s$^{-1}$ cm$^{-2}$) &                 
& (mag) & (erg s$^{-1}$ cm$^{-2}$) & (erg s$^{-1}$)   & 
(10$^{49}$ s$^{-1}$)  \\
\noalign{\smallskip\hrule\smallskip}
\citet{Kennicutt1983}  & 2\phantom{.0} & 7.9$\times$10$^{-12}$             & 
0.05           & 0.00 & 7.6$\times$10$^{-12}$     & 39.73   &  \phantom{0}400  \\
\citet{Lonsdale1987}   & 2\phantom{.0} & 7.9$\times$10$^{-12}$     & 0.05     & 0.43
& 2.1$\times$10$^{-11}$      & 40.17                &  1100 \\
\citet{Helmboldt2004}  & 2:\phantom{0}  & 1.4$\times$10$^{-11}$   &  0.05   &  0.29:
&  2.6$\times 10^{-11}$ & 40.28 & 1410 \\ 
\citet{Hunter2004}     & 2.2  & 1.4$\times$10$^{-11}$ &  0.00    & 0.16
& 2.1$\times 10^{-11}$ & 40.17 & 1110 \\
\citet{Kennicutt2008}  & 2\phantom{.0} & 1.1$\times$10$^{-11}$            & 
0.08           & 0.07
& 1.2$\times$10$^{-11}$      & 39.92                & \phantom{0}615 \\
This study             & 0.2 & 5.1$\times$10$^{-12}$       & 0.03           
& 0.16 & 7.2$\times$10$^{-12}$     & 39.71                &  
\phantom{0}380  \\
\citet{Johnson2003}    &         1:\phantom{0}   &               &                
&                            &          &                      & \phantom{0}530 \\
\noalign{\smallskip\hrule\smallskip}
\multicolumn{8}{l}{$\ast$ Approximate aperture radii.}
\end{tabular}
\label{Ha}
\end{table*}

In summary, we identify the two strong He\,{\sc ii} $\lambda$4686 emission
line sources in IC~4662 with 1--2 young, luminous WN5--6 stars (A1-WR1 and 
A2-WR1), and weak He\,{\sc ii} $\lambda$4686 emission line sources with
either single/binary systems (A1-WR2, A1-WR3/A1-WR4) or ONC-like 
clusters (A2-WR2, A2-WR3). The highest mass, young, optically visible 
cluster IC~4662 is A1-C1, with a probable mass no greater than a factor of 
two higher than Orion. The few point sources in IC~4662 that are visually 
brighter than A1-C1 are probably yellow (AF) supergiants. IC~4662 does 
not  appear to host any  particularly  young, massive, optically visible
star  clusters.

\section{Discussion}

In this section we turn to the  global properties of IC~4662, 
and compare it to other nearby metal-poor star forming galaxies.

\subsection{Embedded clusters in IC~4662?}

In Table~\ref{Ha}, we compare the ACS/HRC F658N-derived number of Lyman 
continuum photons of IC~4662-A with previous ground-based 
H$\alpha$ surveys of the entire galaxy, 
after adjustment to a  common  distance of  2.44~Mpc and including 
corrections for the contribution  of [N\,{\sc ii}] $\lambda$6548, 6584 to 
measured fluxes. 

ACS H$\alpha$  fluxes  are 35--50\% lower than ground-based surveys, 
largely due to its much smaller field-of-view, encompassing just A1 and 
A2. Together, the extended sources B and C (lying to the south west and 
west) and D (to the south east) provide  $\sim$30\% of the total H$\alpha$  
flux of the galaxy, as noted in Section 2, largely 
explaining these differences (recall Fig.~\ref{allregions}).
A1 and A2 each possess similar H$\alpha$ luminosities to the NGC~346
giant H\,{\sc ii} region in the SMC \citep{Relano2002}.

Table~\ref{Ha} also includes thermal 3\,cm continuum-derived ionizing 
fluxes of A1 and A2 from \citet{Johnson2003}, sampling a region of 
IC~4662 that is similar to ACS/HRC. A comparison between these two values 
suggests $\sim$30\% of the ionizing fluxes of A1 and A2 are deeply 
embedded within these regions, in agreement with  \citet{Johnson2003}, 
i.e. arguing in  favour of a number of very young compact 
(ultradense) H\,{\sc ii}  regions. However, \citet{Johnson2003} arrived at 
this conclusion from a comparison between their radio observations and 
optically derived  Lyman continuum fluxes of \citet{Heydari1990}. The latter
was drawn from  \citet{Lonsdale1987}, who adopted the H$\alpha$ flux of 
\citet{Kennicutt1983} and a high, mean interstellar extinction
of $E(B-V)$ = 0.43 mag for their sample of spiral and irregular
galaxies. From Table~\ref{Ha}, as a result of this assumption,
their inferred number of  Lyman continuum photons is a factor of 2.7 
larger than  \citet{Kennicutt1983}, and a factor of two larger than 
\citet{Johnson2003}.

In order to reconcile the optical and radio results there would need
to be ionizing sources contributing to the radio-derived Lyman continuum
flux of IC~4662 A1 and A2 that were undetected at H$\alpha$. 
Young, embedded H\,{\sc ii} regions would be expected to be
bright at both mid-IR (due to warm dust) and radio (free-free emission) wavelengths.
We have therefore inspected 
the Spitzer/IRAC images of IC~4662 from \citet{Gilbert2008} and
note the presence of several IR sources that have no visual counterpart. 
Source N from \citet{Johnson2003} is relatively bright at
3.6\,$\mu$m, as is a source  $\sim$8 arcsec 
south-west of  A2--HII~1, while a fainter 8\,$\mu$m source is seen 
$\sim$10 arcsec south of  A2--HII~1.  Of these, solely IC4662~N represents a 
convincing candidate embedded ionizing source, even though it is 
relatively faint at 3--6\,cm.  Other IRAC candidates lack radio emission, 
while the radio continuum sub-peak north east  of A1--HII~2 has no mid-IR counterpart.

In summary, a comparison between ACS/HRC imaging of IC~4662 and previous
H$\alpha$ and radio surveys suggests 70\% of the total ionizing flux
of the galaxy originates from the giant H\,{\sc ii} regions A1 and A2,
with a non-negligible minority of their Lyman continuum photons 
apparently produced by deeply embedded young massive stars, for which only
IC~4662 N provides a realistic candidate to date, 
from an inspection of Spitzer/IRAC images.
The majority of their ionizing budget is detected
optically, consistent with the age (4.5 Myr) and stellar content (WR 
stars) inferred from VLT/FORS2 spectroscopy. We are unable to reconcile
our results with the very high {\it average} extinction of $A_{\rm V}$ = 
11--25  (9--20) mag for A1 (A2), deduced by \citet{Gilbert2008} from
Spitzer/IRS observations. IRAC sources mentioned above could
suffer such high dust extinctions. However, since the IRS results represent 
averages  across A1 and A2, the low average extinction of $A_{\rm V} \sim 0.5$ mag 
obtained for the, more luminous, optically visible H\,{\sc ii} 
regions would be expected to largely dominate the mean extinction.

\begin{table*}
\caption{Comparison of global properties of IC~4662 with other nearby 
SMC-metallicity star-forming galaxies. Ionizing fluxes are taken
from  extinction  corrected  H$\alpha$ fluxes (SMC, IC4662) or radio 
continuum observations  (IC~4662, IC~10, NGC~1569).}
\label{comparison}
\begin{tabular}{
c@{\hspace{1.5mm}}
c@{\hspace{1.5mm}}
c@{\hspace{1.5mm}}
c@{\hspace{1.5mm}}
c@{\hspace{1.5mm}}
c@{\hspace{1.5mm}}
c@{\hspace{1.5mm}}
c@{\hspace{1.5mm}}
c@{\hspace{1.5mm}}
c@{\hspace{1.5mm}}
c@{\hspace{1.5mm}}
c@{\hspace{1.5mm}}
c@{\hspace{1.5mm}}
c@{\hspace{1.5mm}}c}
\noalign{\hrule\smallskip\hrule\smallskip}
Galaxy & d    & Ref & M$_{B}$ & $R_{25}$ & $R_{D}$ & $\log$(O/H) & Ref & 
$\log Q_{0}$ & Ref        & SFR$^{K98}\ddag$  & SFR$^{C08}\ddag$ & 
SFR$^{L08}\ddag$
& $\Sigma_{\rm SFR}^{R_{25}}\ast$  & $\Sigma_{\rm SFR}^{R_D}\ast$\\
       & (Mpc) & & (mag)   & (kpc)   & (kpc)    &  +12        &     & 
(s$^{-1}$)   &     & 
(M$_{\odot}$yr$^{-1}$)&(M$_{\odot}$yr$^{-1}$)&(M$_{\odot}$yr$^{-1}$)&
(M$_{\odot}$yr$^{-1}$kpc$^{-2}$) &
(M$_{\odot}$yr$^{-1}$kpc$^{-2}$)\\
\noalign{\smallskip\hrule\smallskip}
SMC     & 0.06 & 11& --16.36  & 2.76    & ..      & 8.13        & 1    & 
51.54  & 2 (H$\alpha$)\phantom{0.} &  0.038     & 0.016        & 0.010      & 0.0016 & .. \\
IC~10   &  0.59 & 10 & --15.45  & 0.54    & 0.34    & 8.1\phantom{0}         & 6    & 
51.61      & 7 (radio)     &  0.044     & 0.019        & 0.011      & 0.049\phantom{0}  & 0.12 \\
IC~4662  & 2.44 & 8 &  --15.22 & 0.98    & 0.25    & 8.09        & 3    & 
51.72      & 4 (radio)     &  0.058     & 0.024        & 0.015      & 
0.019\phantom{0}  & 0.3\phantom{0}  \\
        &      & &          &         &         &             &      & 
51.79      & 2 (H$\alpha$)\phantom{0.} &  0.068     & 0.028        & 0.018      & 0.022\phantom{0}  & 0.36 \\ 
NGC~1569& 3.36 & 9 &  --17.98 &  1.77   & 0.38    & 8.19        & 5    & 
52.80      
& 2 (radio)     &  0.674     & 0.287        & 0.175      & 0.07\phantom{00}   & 
1.5\phantom{0}  
\\ 
\noalign{\smallskip\hrule\smallskip}
\multicolumn{15}{l}{\footnotesize{1 \citet{Russell1990}, 2 
\citet{Kennicutt2008}, 3 This work, 4 \citet{Johnson2003}, 5 
\citet{Kobulnicky1997}, 6 \citet{Garnett1990} }} \\
\multicolumn{15}{l}{\footnotesize{7 
\citet{Gregory1996}, 8 \citet{Karachentsev2006}, 
9 \citet{Grocholski2008}, 10 \citet{Borissova2000}, 11 
\citet{Hilditch2005} }} \\
\multicolumn{15}{l}{$\ddag$: SFR prescriptions from K98 
\citep{Kennicutt1998}, C08 \nocite{Conti2008} (Conti  et al. 2008) or
L08 \citep{Leitherer2008}} \\
\multicolumn{15}{l}{$\ast$: $\Sigma_{\rm SFR}$ for $R_{25}$ and $R_{\rm 
D}$ using the  K98 prescription for SFR.}\\
\end{tabular}
\end{table*}

The total number of WR stars in IC~4662-A1 and A2 is probably no greater 
than 10 if A1-WR1 and A2-WR1 indeed host 1--2 luminous WN stars. 
Therefore, N(WR)/N(O)$\sim$0.01, if we were to base
the total number of O stars in A1 and A2 upon the H$\alpha$ luminosity 
from HST/ACS and age inferred from the VLT/FORS2 H$\beta$ equivalent 
widths. These statistics are similar to the SMC, and are consistent with 
the maximum 0.02 ratio  that is  predicted  by SMC-metallicity 
evolutionary models \citep{Meynet&Maeder2005}. 

Of course, it should be bourne in mind that estimates of WR and O star 
populations suffer from significant uncertainties, namely a poorly
defined WN line luminosity for SMC-metallicity stars, plus indirect O
star numbers from nebular methods. If we were to rely solely upon the SMC
metallicity WR calibration of \citet{Crowther&Hadfield2006}, we would 
imply a WR population larger by one order of magnitude. Given the low
statistics of WR stars in the SMC, alternative calibrations should be 
obtained to permit a more  robust line luminosity calibration  for WR 
stars at low metallicity. For example, IC~10 -- a
SMC-metallicity star-forming galaxy within the Local Group -- hosts many
more Wolf-Rayet stars than the SMC \citep{Massey1995, Crowther2003}.

\subsection{IC 4662 in context}

\citet{Hunter2001} and \citet{Hunter2004} have presented a comparison 
between the
properties of IC~4662 and other nearby dwarf irregular galaxies, while
\citet{McQuinn2009} consider the recent star formation history of IC~4662
plus two other dwarf galaxies. In Table~\ref{comparison} we present global 
properties of IC~4662, based in part upon our study, with nearby galaxies 
exhibiting metallicities similar to that of the Small Magellanic Cloud, 
i.e. the SMC itself,
IC~10 and NGC~1569. Of these, solely NGC~1569 is generally recognised as the
nearest bona-fide starburst galaxy. Absolute magnitudes, $M_{\rm B}$
are corrected for foreground reddening and together with isophotal
radii, $R_{25}$ (the $B$-band surface brightness of 25 mag
arcsec$^{-2}$), are drawn from \citet{deVaucouleurs1991}, while
$R_{D}$ was introduced by \citet{Hunter2004} as a scale length from
$V$-band images,\footnote{$R_{D}$ is obtained from a fit to the
$V$-band surface photometry profile, $\mu = \mu_{0} + 1.086 R/R_{D}$
(D.A. Hunter, priv. comm.)} which they considered to be a more meaningful
measure of the galaxy size than $R_{25}$.

The SMC, IC~10 and IC~4662 possess similar Lyman continuum fluxes,
with NGC~1569 an order of magnitude higher. As discussed above,
young ionizing clusters may be deeply embedded, so radio continuum
fluxes were ideally used to derive global ionizing fluxes, with the 
exception of the SMC, for which extinction corrected H$\alpha$ fluxes 
were used \citep{Kennicutt2008}. The usual method of converting H$\alpha$
luminosities or radio continuum fluxes into star formation rates
is \citet{Kennicutt1998}. However, this approach  adopts a
Salpeter IMF. For an alternative \citet{Kroupa2002} IMF, the 
conversion of Conti et al. (2008)\nocite{Conti2008} should be used 
instead, based on \citet{Meynet1994} evolutionary models, 
reducing the inferred star formation rate by $\sim$60\%. Still
lower star formation rates would be implied using contemporary metal-poor
evolutionary models of \citet{Meynet&Maeder2005}, as discussed by 
\citet{Leitherer2008}. 

Results for all three methods are presented in Table~\ref{comparison}, 
together with the star formation rate per unit area, $\Sigma_{\rm SFR}$, 
as derived from the standard \citet{Kennicutt1998} approach using both
$R_{25}$ and $R_{D}$.
Based upon the standard \citet{Kennicutt1998} star formation rate
approach, one definition of a starburst galaxy is a star formation rate 
intensity in excess of $\Sigma_{\rm SFR}^{K98} \sim$ 0.1 
$M_{\odot}$ yr$^{-1}$  kpc$^{-2}$ \citep{Kennicutt2005}.   From this  
definition, using the 
photometric radius $R_{25}$, NGC~1569, with $\Sigma_{\rm SFR}^{K98}  = 
0.07  M_{\odot}$  yr$^{-1}$ kpc$^{-2}$ would narrowly  fail to be 
classified as starburst galaxy,  followed in turn by IC~10 and IC~4662, 
whether using the radio flux -- primarily sampling A1 and A2 -- or the  
H$\alpha$ flux of \citet{Kennicutt2008} for the latter. 

Alternatively, using the definition of $R_{D}$ from \citet{Hunter2004}
both NGC~1569 and IC~4662 would comfortably qualify as starbursts, with
IC~10 a borderline case. \citet{Kennicutt2005} proposed that the 
radii of the H\,{\sc ii} regions within galaxies, $R_{\rm HII}$, is 
preferable to photometric radii when defining surface intensities
-- see also \citet{Martin&Kennicutt2001}. For example, the H$\alpha$ 
(and radio) flux of IC~4662 A1 and A2 can comfortably be accommodated 
within an aperture of radius 10 arcsec ($\sim$120 pc), from which 
a localised star formation intensity of 1.2 $M_{\odot}$ yr$^{-1}$ 
kpc$^{-2}$ would be obtained. Instead, \citet{Hunter2004} quote a
H\,{\sc ii} radius of $R_{\rm HII}$ = 1.14 kpc, as a result of 
region IC~4662--D being located 
far from the main body of the galaxy (recall Fig.~\ref{allregions}). 

\citet{Lee2009} use an integrated H$\alpha$ equivalent
width threshold of 100$\AA$ to identify bursts within dwarf galaxies. This
 corresponds to a stellar birthrate of  $\sim$2.5 -- the current star 
formation rate divided by the recent average  -- on which basis IC~4662 
represents a borderline starburst, with an 
equivalent width of $\sim101\pm10\AA$. Within the local volume (11\,Mpc) 
only a handful of dwarf galaxies fainter than $M_{\rm B}\sim -17$ 
mag would be characterised as starbursts on  this basis, including 
NGC~1569 and other well known cases NGC~1705, NGC~2366, NGC~3125 and 
NGC~5253.

\section{Conclusions}

We have carried out a ground-based VLT/FORS2 imaging and spectroscopic
survey of the massive stellar content of giant H\,{\sc ii} regions A1 and 
A2 within the nearby star forming galaxy IC~4662, supplemented by archival
HST/ACS imaging. 

From a comparison between radio continuum observations of IC~4662 with
HST/ACS H$\alpha$ imaging, we find that a minority of the ionizing fluxes 
of A1 and A2 likely result from  deeply  embedded clusters,  in agreement 
with  conclusions of \citet{Johnson2003}. A1 and A2 each possess similar 
H$\alpha$ line luminosities to NGC~346 in the SMC.

We identify several sources hosting Wolf-Rayet stars (or nebular 
He\,{\sc ii} $\lambda$4686 emission). There is one bright and  one faint 
source within each of A1 and A2 plus one faint region $\sim$8 arcsec 
north west of A1 that we label A1-NW. 
Spectroscopy reveals  that mid-type WN  stars are the dominate subtypes 
within A1 and A2, in addition to one WC star in each of A2 and A1-NW 
together with  nebular  He\,{\sc ii}  $\lambda$4686 emission in two 
instances. ACS/HRC  imaging  permits  individual sources responsible for 
emission line features to be  identified, which we conclude are a mixture 
of luminous WN stars (or WR binaries) and $\sim$5 Myr old clusters, with 
masses up to two times larger than the Orion Nebula Cluster. In contrast
with other nearby dwarf irregular galaxies exhibiting starburst activity,
IC~4662 appears to lack young, optically bright very massive clusters. For 
example, the  brightest young cluster of IC~4662 is A1-WR2 for which 
$M_{\rm F550M} =  -9$ mag, versus $M_{\rm F555W} = -11.9$ mag for the 
5~Myr old cluster NGC~1569~A2 \citep{Maoz2001, Sirianni2005}.  

Finally, we compare the global properties of IC~4662 with other nearby
SMC-metallicity star forming galaxies. Based upon photometric 
radius $R_{25}$, IC~4662 is not a starburst galaxy, even though its global 
star  formation  intensity is an order of magnitude higher than that of 
the SMC.  NGC~1569  and IC~10 possess star formation intensities a factor 
of  $\sim$3 times higher than IC~4662, the former generally recognised 
as  the nearest bona-fide starburst galaxy. If we were to follow the
approach of \citet{Hunter2004}, by defining an alternative scale 
length, $R_{D}$, measured from $V$-band images to define the size of the 
galaxy, both NGC~1569 and IC~4662 would readily qualify as starbursts
-- see also \citet{Lee2009}.

Wolf-Rayet stars within low metallicity irregular galaxies such 
as IC~4662 are leading  candidates  for Type Ic supernova-GRBs 
\citep{Fruchter2006, Hammer2006, Modjaz2008}. Establishing 
the spatial location of WR stars within such environments would permit the 
progenitors of such explosions, in the case of a core-collapse supernova 
fortuitously occuring in the near future within IC~4662.

\section{Acknowledgements} 
Thanks to Hans Schild for co-observing IC~4662,
Janice Lee for permitting use of unpublished CTIO H$\alpha$ 
images of IC~4662, Deidre Hunter for details of her H$\alpha$ observations, and to 
Bill Vacca for useful discussions. JLB acknowledges financial support from 
STFC. This research has made extensive use of the NASA/IPAC Extragalactic 
Database (NED) which is operated by the Jet Propulsion Laboratory, 
California Institute of Technology, under contract with the National 
Aeronautics and Space Administration.

\setlength{\bibsep}{0pt}

\bibliographystyle{aa}

\label{lastpage}

\end{document}